# Diffusion-weighted MRI – guided needle biopsies permit quantitative tumor heterogeneity assessment and cell load estimation

Yi Yin[a,b,§,*], Kai Breuhahn[c], Hans-Ulrich Kauczor[d,e,f,g], Oliver Sedlaczek[d,e,f,g,#], Irene E. Vignon-Clementel[a,b,#], Dirk Drasdo[a,b,#,*]

[a] *Inria, Centre de recherche de Paris, 75012 Paris, France*
[b] *Laboratoire Jacques-Louis Lions, Sorbonne Université UPMC, 75252 Paris, France*
[c] *Institute of Pathology, University Hospital Heidelberg, 69120 Heidelberg, Germany*
[d] *Translational Lung Research Center Heidelberg (TLRC), member of the German Center for Lung Research (DZL), 69120 Heidelberg, Germany*
[e] *Department of Diagnostic and Interventional Radiology, University Hospital of Heidelberg, 69120 Heidelberg, Germany*
[f] *Department of Diagnostic and Interventional Radiology with Nuclear Medicine, Thoraxklinik at University of Heidelberg, 69126 Heidelberg, Germany*
[g] *Department of Diagnostic Radiology, DKFZ, 69120 Heidelberg, Germany*
[§] *Current affiliation: Nuffield Department of Women's & Reproductive Health, University of Oxford, Oxford, OX3 9DU, United Kingdom*

[#] *Shared senior authorship*
[*] *corresponding authors*

**Abstract**

Motivation: Quantitative information on tumor heterogeneity and cell load could assist in designing effective and refined personalized treatment strategies. It was recently shown by us that such information can be inferred from the diffusion parameter *D* derived from the diffusion-weighted MRI (DWI) if a relation between *D* and cell density can be established. However, such relation cannot a priori be assumed to be constant for all patients and tumor types. Hence to assist in clinical decisions in palliative settings, the relation needs to be established without tumor resection. It is here demonstrated that biopsies may contain sufficient information for this purpose if the localization of biopsies is chosen as systematically elaborated in this paper.

Methods: The cell density is analyzed in virtual biopsies from the histological tissue images of the same solid tumor for which previously large tissue samples are studied. This permits studying the impact of biopsy location and validating our findings in virtual biopsies by direct comparisons with large tumor samples. A superpixel-based method for automated optimal localization of biopsies from the DWI *D*-map is proposed. The performance of the DWI-guided procedure is evaluated by extensive simulations of biopsy needle paths. The method is by construction patient-specific, not requiring prior calibration curves to relate DW-images to cell density.

Results: Needle biopsies yield sufficient histological information to establish a quantitative relationship between *D*-value and cell density, and to investigate tumor cellularity and heterogeneity, provided they are taken from regions with high, intermediate, and low *D*-value in DWI. The automated localization of the biopsy regions is demonstrated from a NSCLC (non-small cell lung cancer) patient tumor. In this case, even two or three biopsies give a reasonable estimate. Simulations of needle biopsies under different conditions indicate that the DWI-guidance highly improves the estimation results.

Conclusion: Tumor cellularity and heterogeneity in solid tumors may be reliably investigated from DWI and a few needle biopsies that are sampled in regions of well-separated *D*-values, excluding adipose tissue. This procedure could provide a way of embedding in the clinical workflow assistance in cancer diagnosis and treatment based on personalized information.

---

**1. Introduction**

Cancer cell heterogeneity is regularly observed in tumors, which increases the complexity of cancer diagnosis and personalized treatment. Tumor heterogeneity can be assessed by histological analysis after resection (Potts et al., 2012; Mani et al., 2016) and diagnostic imaging, such as magnetic resonance imaging (MRI) (Just, 2014; Irving et al., 2016; Reizine et al., 2020), ultrasound (Al-Kadi et al., 2015), computed tomography (CT) (Ganeshan and Miles, 2013), and positron emission tomography (PET) (Vriens et al., 2012; Soussan et al., 2014). DWI is a MR-measurement reflecting basically the mobility of free water molecules within a tissue volume at different scales, thus indirectly providing information on tissue cellularity (Bihan et al., 1986). In general, a tissue of highly packed cells exhibits a low diffusion coefficient. Heterogeneous diffusion could thus be considered as one of the indicators of tumor heterogeneity (Asselin, 2012; Just, 2014; Ryu et al., 2014; Fan et al., 2017). The advantageous aspects of these imaging techniques include their non-invasiveness and their representation of whole tumor compared to the resection analysis. However, non-invasive image modalities provide indirect information of tumor heterogeneity at a spatial resolution much coarser than the cellular level.

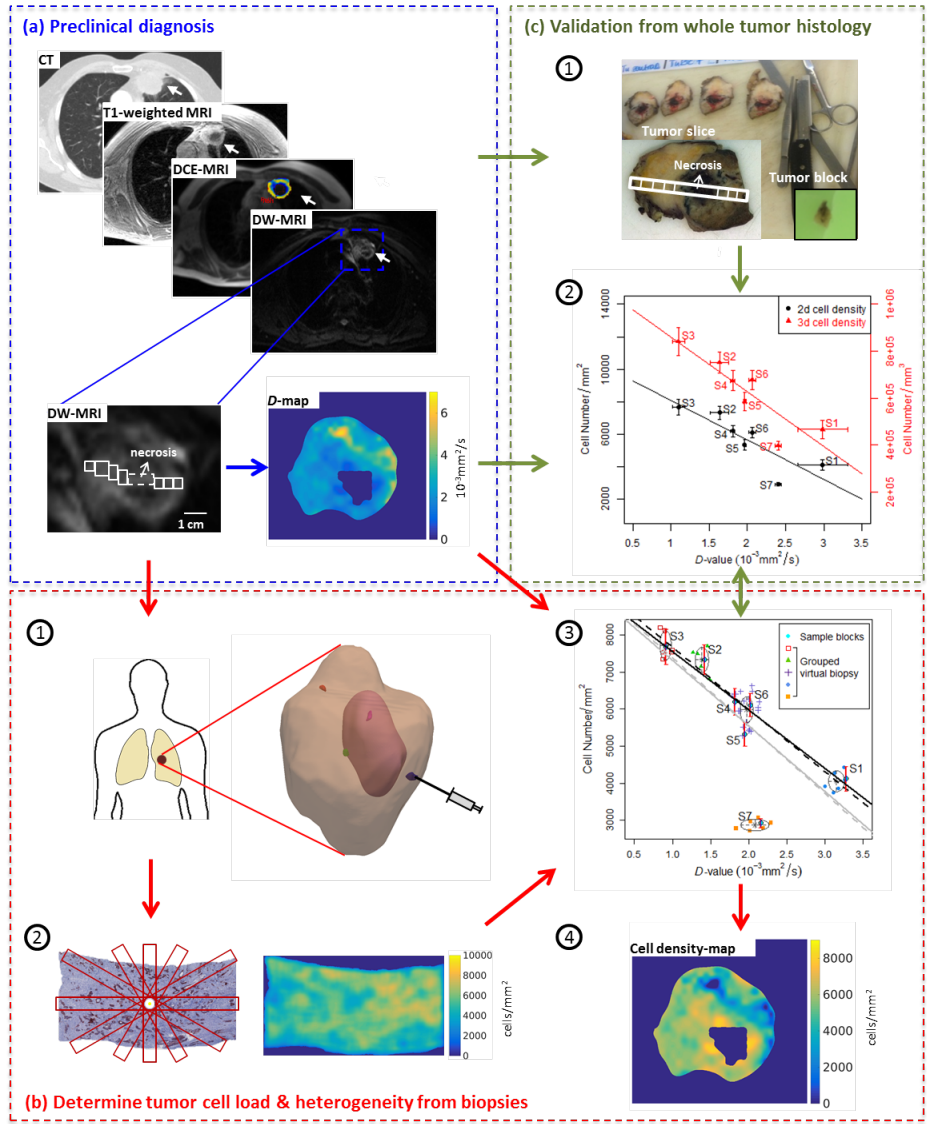

**Fig. 1.** Proof of concept for DWI-guided biopsy. (a) Preclinical diagnosis based on multimodalities (CT, T1-weighted MRI, DCE-MRI, and DW-MRI) for tumor localization and classification. The parametric *D*-map is generated by estimating the voxel-wise diffusion coefficient from DWI data. (b) Tumor cell load and heterogeneity estimation from biopsies. Candidate biopsy regions are shown inside the tumor volume of the studied NSCLC patient tumor outside of the pink necrotic region (1). A negative linear relationship (3) is established between the cell density and the DWI *D*-value from virtual biopsies (red rectangles on the bottom left (2)), which is a necessary step to produce the cell density map to quantify tumor heterogeneity (4). (c) Validation from resected large tumor samples that show negative correlations between the DWI *D*-value and both 2d (black) and 3d (red) cell densities as discussed in (Yin et al., 2018).

Evaluation of micro-architecture and structural components (e.g. DNA, proteins, and organelles) of cells and tissues is often performed based on small tissue fragments sampled through CT-/MR-/Ultrasound-guided needle biopsy. Biopsies are commonly used in the diagnosis and staging of cancers. As a single biopsy sample collected at one tumor location is in many cases insufficient for diagnosis (Bedard et al., 2013), samples from multiple regions are probed to provide sufficient information for therapy decisions (Chen et al., 2013; Calhoun and Anderson, 2014). A greater number of probed locations reduces the sampling error (Dekker et al., 2013). The suggested number of biopsies taken varies from two in breast cancer diagnosis to 12 in cancer stratification in the blind probing for prostate cancer (de Lucena et al., 2007; Shure and Astarita, 1983; Ofiara et al., 2012; Bjurlin and Taneja, 2014). For lung cancer, 4 endobronchial biopsy specimens were supposed to be adequate for optimal diagnosis of endobronchial tumor in central lesions (Shure and Astarita, 1983), while in order to gather enough tissue to perform a more detailed molecular analysis, Ofiara et al. suggested to obtain up to 6 specimens (Ofiara et al., 2012).

Different strategies have been proposed to take accurate and representative biopsies. Kryvenko et al. defined rules of biopsy to predict insignificant prostate cancer at radical prostatectomy (Kryvenko et al., 2014). In (Finn, 2016), Finn introduced the roles of liver biopsies for guiding or stratifying treatment of hepatocellular carcinoma. Kumaraswamy and Carder established the number of histological levels necessary for the evaluation of needle core biopsy specimens taken from areas of mammographic calcification in breast cancer (Kumaraswamy and Carder, 2007).

The aforementioned strategies mainly target tumor and cell phenotyping. However, to the best of our knowledge, biopsy criteria (the optimal number and location) in solid tumor enabling quantitative estimation of the whole tumor cellularity and heterogeneity have not yet been addressed.

Biopsies can be taken under MRI, ultrasound, CT, or x-ray guidance, depending on the imaging modality that permits the best biopsy guidance (Zhou et al., 2014). Nowadays, the interest in MRI-guided biopsy is increasing (Tse et al., 2009; Seifabadi et al., 2012, 2016; Li et al., 2017; Henken et al., 2017), because MRI provides excellent soft tissue contrast and does not use ionizing radiation (Henken et al., 2017). In addition, MRI can provide structural and functional tissue parameters such as diffusion coefficient from DWI, aiding in cancer diagnosis and treatment assessment (Malayeri et al., 2011). In fact, MR-guidance plays an increasing role in choosing the significant biopsies. While biopsies are only taken very rarely in the MR-theatre (i.e. mainly in very selected cases of breast cancer), typical examples for fusion techniques are the prostate biopsies, fusing MR-images with Ultrasound during the acquisition of the biopsies increasing the detection rate from 60-80% to 97% (Radtke et al., 2016).

In this study, optimal biopsy criteria are investigated and a protocol is proposed for performing needle biopsies to quantify the cellularity of the whole tumor under the guidance of DWI, as listed below:

---
**Protocol: DWI-guided biopsy**

---

1. **Perform DW-MRI** (Section 2.2)
2. **Generate $D$-maps** for MRI slices containing tumor tissue and excluding fat tissue based on the observation from T1/T1-fat-sat MRI (Section 3.3)
3. **Interpolate $D$-maps** to assign appropriate $D$-values to the intermediate positions between the centers of voxels (Section 3.3.1)
4. **Identify the possible locations of biopsies** based on superpixel partition of well-separated $D$-value regions (Section 3.4)
5. **Take biopsies** from candidate regions, which are characterized by at least three different $D$-values (Section 3.5 & 3.6)
6. **Perform histological analysis** of the biopsies (Section 3.2)
7. **Draw the cell density versus the $D$-value** from these biopsies and infer its relationship from the plot (Section 4.3)
8. **Estimate the cellularity and heterogeneity** in the entire tumor based on the $D$-value vs. cell density relationship (Section 4.4)
9. **Estimate the cancer cell fraction** from the positive correlation between the cancer and the total cell (cancer and non-cancer) densities (Section 4.4)

---

The applicability of the proposed protocol is demonstrated by using the histological material and DWI data from the same NSCLC (non-small cell lung cancer) tumor (Fig. 1). Multi-modalities-imaging had been performed prior to surgical resection of the tumor (Fig. 1(a)). As explained in a previous communication (Yin et al., 2018), part of the resected tumor was stained and cut into 7 tissue blocks for histological analysis. The 7 blocks were co-registered to DWI to relate the cell density $\rho$ of the 7 blocks with the corresponding $D$-value from the DWI (Fig. 1(c), (Yin et al., 2018)). However, such an analysis as shown in Fig. 1(c) would only be possible after surgery and therefore not be applicable for diagnosis or therapy planning. One cannot a priori assume that for each tumor and tumor subtype the same relation between *D*-value and cell density applies, so the relation $\rho(D)$ established in (Yin et al., 2018) cannot a priori be readily used. Hence, it is here investigated if an equivalent information as from the tumor blocks could also be obtained from several biopsies (Fig. 1(b)) as in this case, this information would be patient-specific. For this purpose, virtual biopsies are defined by drawing small ROIs (regions of interest) in the resected large tumor samples as shown in Fig. 1(b-2) and Fig. 2. Each ROI is rectangle and contains as much tissue as a biopsy needle might capture. The corresponding regions of virtual biopsies are then identified in the *D*-map generated from the DWI. As the biopsy tissue samples are smaller in width than the voxel dimension of DWI/D-map, the *D*-map to find sub-voxel location of ROI of virtual biopsies is interpolated and the corresponding *D*-values are calculated based on the interpolated voxels within the ROI. If the (virtual) biopsies are chosen from regions of at least one high, low and intermediate *D*-value (i.e., at least three different *D*-values) excluding fat tissue, a negative linear relationship is obtained between the biopsy cell density and the *D*-value showing a similar slope as that obtained from the large tumor block samples discussed in ref. (Yin et al., 2018) (Fig. 1(b-3)). Knowing this relationship extracted from the virtual biopsy analysis, as well as the *D*-value of each tumor voxel in DWI data, the corresponding cell densities of each voxel can be computed, which then reflects the tumor heterogeneity (Fig. 1(b-4)).

For automated localization of the representative biopsies, a superpixel-based method is proposed to find the candidate biopsy regions with representative *D*-values. In the successive step the biopsy procedure is simulated to explain how to perform biopsies to reach these regions, whereby the previously (Yin et. al. 2018) studied lung tumor serves as an example. Sets of needle paths (in Fig. 6 referred to as "optimal") are probed to ensure the candidate regions on the *D*-map are captured by iteratively examining all the possible paths (needle access points, needle depth and orientation inside the tumor) respect the constraints encountered in practice (i.e. proximity of vital organs, vessels, pleural-space passages, etc.). The *D*-value of biopsy is computed as the average *D*-value within the rectangle ROI at the tip of each path on the *D*-map, and the corresponding cell density is estimated from the tissue material taken by the biopsy needle. Based on these values a relation between *D*-value and cell density is established. Besides, the DWI-guided biopsies with the non-guided biopsies with and without constraints are compared by simulations to evaluate the potential benefit of a DWI guidance.

In summary, study of this NSCLC tissue indicates that the analysis of large tumor blocks could be replaced by the analysis of only three to four biopsies in properly chosen regions, arriving at almost the same quantitative estimate of tumor heterogeneity and cellularity. These regions can be identified from the *D*-map derived from the DW-images. Even 2 biopsies may give a reasonable estimate. The use of biopsies makes the procedure of quantitative assessment of the whole tumor applicable in the clinics. 3 biopsies are nevertheless recommended to obtain a "red flag" in case one biopsy would be defective. Moreover, it is demonstrated that biopsies even permit to establish a relationship between cell density (including both cancer and non-cancer cells) and cancer cell density such that not only the cell density, but also the cancer cell density could be inferred.

**2. Data description**

*2.1. Patient tumor and histology material*

The primary tumor tissue comes from a 72-year-old patient with a non-small cell lung cancer (NSCLC) who underwent an upper left lobe en-bloc resection at the Department of Thoracic Surgery, Thoraxklinik Heidelberg/Rohrbach (Germany). Informed consent was obtained from the patient. Different imaging modalities were performed prior to surgery, including CT and functional MRIs (T1-weighted, diffusion-weighted, and DCE-MRI) (Fig. 1(a)), to localize the tumor and classify the tumor tissue. In this study, the tumor contour is manually segmented from T1-weighted images, based on the contrast differences. The central necrosis region is identified from DCE-MRI based on the perfusion kinetics (the central necrotic area of the tumor being not perfused).

After resection, the patient was treated with adjuvant radiation therapy. The patient tumor of size 6.5 *cm* in diameter was cut into slices in axial plane identical to the relevant MRIs (Fig. 1(c-1)). The slice above the central cutting line was selected for further analysis. Upon fixation in 4% buffered formalin overnight, the selected slice was cut into tissue blocks of size of 1 *cm* × 1 *cm* × 0.5 *cm* visually matched to the structures identified in the co-registered MRI and CT. The tumor blocks were then transferred in paraffin and cut into thin slide of thickness of 1 – 2 *μm* by an automatic microtome (HM 355S, Thermo Scientific, Braunschweig, Germany). These tissue slides were stained using H&E (hematoxylin/eosin) and AFOG (acid fuchsin orange G), as well as an anti-pan cytokeratin antibody (clone KL1, Abcam, Cambridge, UK) for epitope-specific stain. The slides were then scanned by a Hamamatsu NanoZoomer (Hamamatsu Photonics, Hamamatsu, Japan), providing digitized 2d bright field images (e.g. picture on the bottom left of Fig. 1(b-2)).

*2.2. DWI and diffusion coefficient*

A clinical standard thorax MRI (Puderbach et al., 2007) was performed on a 1.5 T machine (Avanto, Siemens, Erlangen). A navigated diffusion echo-planar imaging sequence was applied with the following parameters: TE 75 *ms*, TR 8364 *ms*, SL 6 *mm*, acquiring 7 *b*-values = {0, 50, 100, 150, 200, 400, and 800 *s/mm²*}. These images have a voxel size of 2.1 *mm* × 2.1 *mm* × 6 *mm*. The DWI axial scan of the patient consists of 40 slices, among which the 9 slices in the middle cover the whole tumor.

In DWI, the signal attenuation can be related to the motion of water in tissue using the following bi-exponential model, called intra-voxel incoherent motion (IVIM) model (Bihan et al., 1986).

$$\frac{S(b)}{S(0)} = (1-f)e^{-bD} + fe^{-b(D+D^*)}, \qquad (1)$$

where $S(b)$ and $S(0)$ are echo signal amplitudes of diffusion-weighted and non-diffusion-weighted intensities, respectively. Parameter $b$ defines the degree of diffusion sensitization depending on the magnitude, duration of the gradient pulse and the time interval between two successive pulses. $D$ is the diffusion coefficient characterizing the restricted mobility of water molecules. It is supposed to be associated with the cell density in the solid portion of the vital part of a tumor. $D^* \gg D$ is the pseudo-diffusion coefficient related to blood flow in the capillary network. $f$ is proportional to the blood vessel volume fraction (Bihan et al., 1988).

In this study, the open-source software MITK Diffusion (Fritzsche et al., 2012) is used to estimate the diffusion coefficient (*D*-value) by fitting the signal attenuation model shown in Eq. (1) to the data, namely the voxel signal versus *b*. All the *b*-values are taken into account in the calculation. A parametric *D*-map of the tumor (Fig. 1(a)) is created by estimation of the *D*-value of each voxel within the tumor region.

**3. Methodology**

*3.1. Virtual biopsy*

The 2d histological images of the tissue slides cut from the tumor blocks as described in Section 2.1 can provide detailed microstructure information. Its size of 1 *cm* × 1 *cm* is large enough for accurate localization of its corresponding region on the DWI images. In routine cancer diagnosis, the needle biopsies are much smaller in size, e.g. the widely used 18G core needle biopsy (Zhou et al., 2014; McCormack et al., 2012) has a nominal inner diameter of 0.84 *mm*. Tissue material taken by a needle is cylinder-shaped, from which the small pieces of tissue sections are cut for pathological exam (Fig. 2(a)). In order to mimic the biopsy procedure, small tissue regions are selected from the middle section of the large tumor block, which are comparable in size and shape with the real biopsies, and thus called here *virtual biopsy*. The ROI of a virtual biopsy is chosen by a rectangle of width 0.84 *mm* and length 4.35 ± 1.35 *mm* (Fig. 2(b)). For each large tumor block, 6 virtual biopsies are taken, which are centered in the centroid of the middle tissue section and rotated by interval of $30^o$.

The ROIs of virtual biopsies in DWI are identified accordingly, which will be further discussed in Section 3.3.

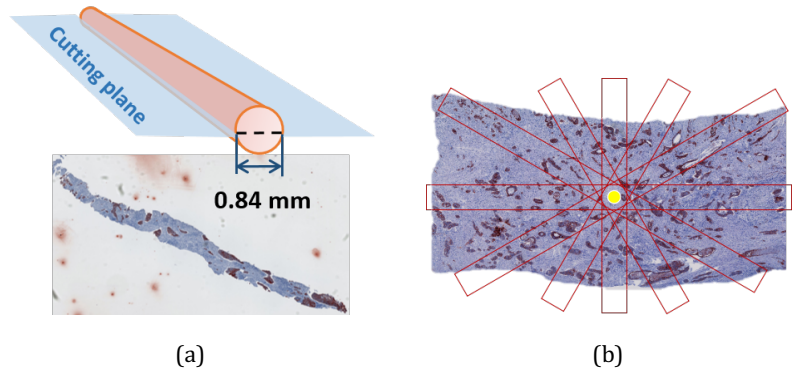

(a)            (b)

**Fig. 2.** Illustration of the real biopsy material in (a) and the virtual biopsy in (b): (a) a piece of tissue section cut from the middle of a cylinder-shaped tissue sample taken by a 18G biopsy needle (BARD®, MAGNUM® Biopsy MN18130); (b) a histological section taken from the large tumor block in which the red rectangles show the virtual biopsy regions and orientations, and the yellow point indicates the tissue centroid.

*3.2. Cell density estimation*

Histology slide imaging provides the high-resolution color images in 2d. An in-house image analysis tool was developed to estimate the 2d and 3d cell densities from the histological images (Yin et al., 2018), which is included as a module in the software TiQuant (Friebel et al., 2015). The Matlab codes can be accessed at Gitlab, https://gitlab.inria.fr/multicellular_modelling/public/cellsegmquant. The tool consists of three processing steps (Fig. 3): (i) pre-processing (image denoising and smoothing), (ii) 2d cell density estimation based on cell nuclei segmentation and classification by using the hematoxylin (nucleus staining) and KL1 (cancerous sensitive region staining) channels of the histological image which are separated by color deconvolution (Macenko et al., 2009), and (iii) 3d cell density estimation by an iterative model-based optimization method using 2d cell density and 2d cell nucleus features. The latter includes nucleus size distribution. Cell nuclei in 3d are approximated by ellipsoids. For the calculation of cell density each cell is assumed to have only one nucleus. The polynuclear cells could be taken into account by membrane staining. A linear relationship between 2d and 3d cell densities is obtained for each cell type, $\rho_{3d}^{\xi} = m^{\xi} \rho_{2d}^{\xi}$ with $\xi \in \{c, nc\}$ for cancer and non-cancer cells, respectively (Yin et al., 2018). These $m^{\xi}$ values have to be determined for each patient based on the 2d and 3d cell densities estimated from the histological images generated from patient's tumor tissue data. Within the DWI-guided biopsy protocol, this data can be gathered by performing biopsies. One way to do this is to determine the 2d cell density and use the algorithm of Yin et. al. (2018) to obtain the 3d cell density.

The histological data of the large tumor blocks are analyzed with the help of the tool, achieving high detection accuracy, namely, up to 0.95 for cell segmentation and 0.96 for cancer/non-cancer cell classification as evaluated by a pathologist (Yin et al., 2018). For each virtual biopsy taken from these large tumor blocks as defined in Section 3.1, the 2d cell density is calculated by dividing the number of detected cell nuclei by the tissue area within the biopsy ROI. The 3d cell density of the tumor block is then obtained based on the linear mapping from 2d cell density to 3d cell densities, using $m^c$ = 96/mm for cancer cells, and $m^{nc}$ = 115/mm for non-cancer cells. Here $m^c$ and $m^{nc}$ were set to the values estimated from the large tumor blocks, as the virtual biopsies are from the same patient tumor regions as these tumor blocks. However, it is equally possible to scan the biopsy tissue to generate digitized histological images. Based on these images, the 2d and 3d cell densities can be computed using the algorithm in (Yin et al., 2018), and finally estimate $m^c$ and $m^{nc}$.

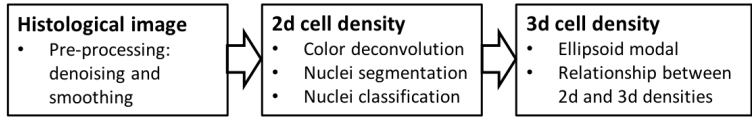

**Fig. 3.** Automated histology analysis: cell segmentation, cancer cell/non-cancer cell classification, cell density estimation in 2d and 3d. (Details see (Yin et al., 2018).)

*3.3. Diffusion coefficient of biopsies*

A parametric *D*-map is generated from the DWI slice (Fig. 4(a)) as described in Section 2.2. It has the same voxel resolution 2.1 *mm* × 2.1 *mm* × 6 *mm* as in DWI. The DWI slice used for the virtual biopsies analysis is the one matched to the tumor slice from which the large tumor blocks were taken to ensure that the DWI data information can be related to the histological information of the same tissue samples (see Fig. 1(a) and (c)). Since the voxel size in each of the three dimensions in DWI *D*-map is larger than the width of a biopsy, the *D*-map is interpolated in order to assign appropriate *D*-values to the intermediate positions between the voxel centers. This facilitates the localization of the accurate boundaries of biopsies. The ROIs of virtual biopsies in the DWI *D*-map can be found, knowing how they are defined in a tumor block and based on the region correspondence of tumor blocks in DWI. The ROIs in DWI have a width of 0.84 *mm* and a length of 4.55 ± 1.45 *mm*, which is similar to the values of ROIs in the histological images.

*3.3.1. D-map interpolation*

The *D*-map of the selected tumor slice is spatially interpolated by the Bicubic method (Burger and Burge, 2008) to permit better identification of the *D*-values that would belong to the region of the biopsy based on the assumption that the transition between voxels with different *D*-values is smooth (Fig. 4(b)). To study the influence of interpolation on the *D*-value distribution, the Kullback-Leibler divergence (MacKay, 2002; also, relative entropy) is calculated between the *D*-value distributions before and after interpolation. This is a measure of the distance between two distributions, which is defined as,

$$Div_{(KL)}(P||Q) = \sum_{l=1}^{N_{bin}} P(l) * log_2\left(\frac{P(l)}{Q(l)}\right), \qquad (2)$$

where *Q* denotes the original distribution and *P* the distribution after interpolation. $N_{bin}$ is the number of bins in the histogram of *D*-values before interpolation. A scaling factor *U* is specified to resize an image by interpolation, e.g. resizing an image of 2×2×2 to 2*U*×2*U*×2*U*. *U* reflects the number of points inserted between the center-points of two neighboring voxels on *D*-map in each dimension. For the studied tumor *U* is set such that this divergence measure converges. The results will be discussed in details in Section 4.2. The ROIs of large tumor samples (Fig. 1(a)) are transferred from the original DWI section to the interpolated *D*-map. ROIs of virtual biopsies on the *D*-map are then localized inside the ROIs of the large tumor samples, having the same position and orientation as the ROIs of biopsies in the tissue section image of the large tumor sample.

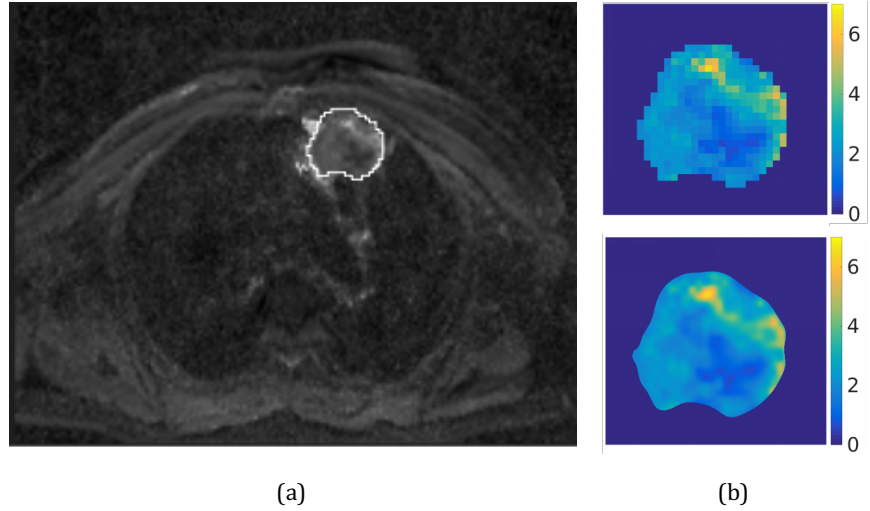

(a)                (b)

**Fig. 4.** $D$-map interpolation. (a) $D$-map obtained from the DWI slice corresponding to the middle tumor plane. The white line delineates the tumor contour; (b) comparison between the $D$-maps ($10^{-3} mm^2/s$) before (top) and after interpolation (bottom; scaling factor $U$ = 20).

### 3.3.2. D-value estimation from ROIs on D-map

The $D$-value of an ROI is usually estimated as the mean value over the ROI (Kono et al., 2001; Hayashida et al., 2006; Ginat et al., 2012),

$$D_{ROI} = \frac{1}{N_{ROI}} \sum_{i=1}^{N_{ROI}} D_i. \tag{3}$$

Here, $D_i$ is the $D$-value of voxel $i$ inside the ROI on $D$-map, and $N_{ROI}$ is the number of voxels within the ROI. The same relation in Eq. (3) is used to compute the $D_{ROI}$. The parameters found agree well with physical reasoning and histological observation based on the arguments detailed in Appendix A.

### 3.4. Superpixel-based automated localization of biopsies

In this study, the biopsies are sought to be performed following the rules listed below:
- Be representative of the whole range of tumor cell density within the tumor
- Have low variance in cell density within the ROI of biopsy (which can be guaranteed by applying superpixel partitions as explained below)
- Be neither too close to tumor boundary (> 0.25 cm), nor be in the necrotic zone
- Exclude fat tissue

A superpixel-based approach is proposed to find the regions fulfilling these rules to perform biopsies from the $D$-map (Fig. 5). Superpixel is a recently-made popular concept in computer vision applications, which partitions an image into small segments by grouping adjacent pixels with similar features (Ren and Malik, 2003). On the *D*-map, adjacent pixels with similar features (here: intensities) that have low variance are grouped together, forming superpixels (Fig. 11(b)). The size of a superpixel is defined to be at least as big as, ideally bigger than, the real size of needle biopsies.

The candidate biopsy regions are selected from these superpixels as the ones whose *D*-values distribute uniformly to cover the range of *D*-values of the whole tumor. If the distribution of D-values is symmetric, performing only one biopsy, $N_{biopsy}$ = 1, the optimum is defined by $D^{OPT}=<D>$. If $N_{biopsy} >1$, for $j$ = 1, …, $N_{biopsy}$,

$$D_j^{OPT} = D_{min} + (j-1)\frac{D_{max}-D_{min}}{N_{biopsy}-1}. \quad (4)$$

In case the *D*-value distribution $Q(D)$ is largely asymmetric with a long tail, the setting $D^{OPT}= D_{Md}$, the median of *D*-values, for $N_{biopsy}$ = 1 may be more appropriate. Equivalently, if $N_{biopsy} >1$, for $j$ = 1, …, $N_{biopsy}$,

$$D_j^{OPT} = D_{ini} + (j-1)\,\Delta_D, \quad (5)$$

where $D_{ini} = D_{Md} - min(|D_{Md} - D_{min}|, |D_{Md} - D_{max}|)$, and $\Delta_D = 2(D_{Md} - D_{ini})/(N_{biopsy} -1)$, as it gives less weight to the part of the distribution that has the longer tail. In order to obtain robust measures, $D_{min}$ and $D_{max}$ are defined as the 2*nd* and 98*th* percentiles of the *D*-values respectively. In case of a symmetric distribution, the formulas yield the same result. Those superpixels in the *D*-map whose average values are closest to $D_j^{OPT}$ are finally selected as the candidate regions for the biopsy.

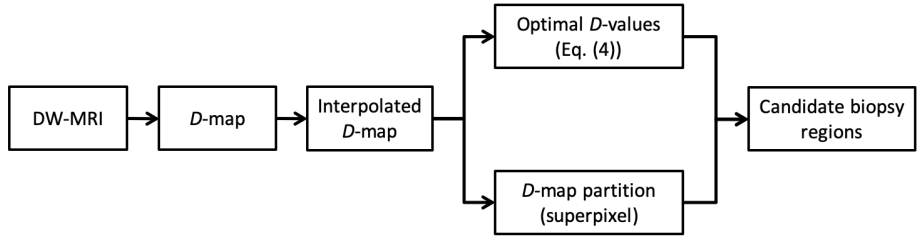

**Fig. 5.** Flowchart of the superpixel-based automated localization of needle biopsies.

### 3.5. Biopsy needle simulation

In the real biopsy procedure, there are always constraints of needles, e.g. the length of a needle and orientation of entrance. In order to simulate a realistic biopsy procedure together with their common constraints in the studied patient lung tumor the rules for performing biopsies (see Fig. 6(a)), adopted by the Univ.

Clinics of Heidelberg are mimicked. For each biopsy procedure with multiple needle paths, these constraints are:
- All the biopsy needle paths have the same access point at the border of the tumor
- The needle paths are within an angle of 20º
- The maximal needle depth inside the tumor is 2.2 cm.

Fig. 6(b) and (c) give two examples of biopsies showing the needle paths of different access points, needle penetration depths and orientations.

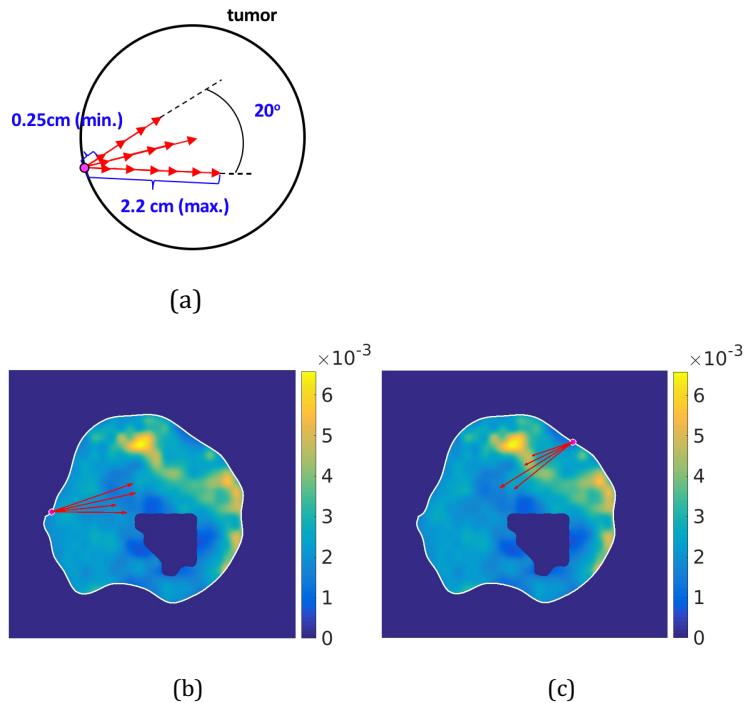

(a)

(b) (c)

**Fig. 6.** Simulation of a biopsy procedure. (a) Illustration of the constraints: the pink point indicates the access point and the red lines with arrows indicate a set of needle paths showing various depths and orientations; (b) and (c) show two examples of the mimicked needle biopsy paths (4 red arrows) on the *D*-map (color code in $mm^2/s$) of the tumor (the interior dark blue area with formally *D=0* being associated with necrosis).

*3.6. Automated guidance of the biopsy*

In a next step, all the possible sets of biopsy needle paths are evaluated under the above-defined constraints to probe the optimal punching point, needle depths and orientations. The optimal set of paths is supposed to reach the candidate biopsy regions as discussed in Section 3.4.

The algorithm iterates through all the access points at the tumor border in clockwise order starting from the 9 o'clock position. The distance between two adjacent access points is chosen of 1 mm. For each access point, a set of needles is mimicked by line segments from the access point to the tumor interior. The access point is stepwise rotated in clockwise order by an increment of 5°, ensuring the needle tips being within the tumor region. For a given orientation, the depths of the needle paths change respectively ranging from 0.25 to 2.2 cm by a step of 0.25 cm as shown in Fig. 6(a). Those sets of needle paths that reach the candidate biopsy regions are picked as the optimal sets for a biopsy. The $D$-values of the tissue material sampled by the optimal biopsy are representative and closest to the optimal $D$-values computed by Eq. (4).

## 4. Results

*4.1. Cell density evaluation of virtual biopsies*

Virtual biopsies are taken from the large tumor blocks of the studied NSCLC patient tumor as described in Section 3.1 (see Fig. 2(b)). The average cell densities obtained from 6 virtual biopsies and the cell density of the same tumor block are pairwise compared as listed in Table 1. The estimation results from local biopsy regions represent the cell density distribution in the large tumor blocks.

**Table 1**

2d cell densities (*mean* ± SD) ($10^3$ *cells/mm$^2$*) of each tumor block and the corresponding virtual biopsies.

| ID of block | Whole tumor block | 6 biopsies per block |
|:---:|:---:|:---:|
| S1 | 4.12 ± 0.34 | 4.05 ± 0.26 |
| S2 | 7.34 ± 0.40 | 7.33 ± 0.32 |
| S3 | 7.68 ± 0.48 | 7.72 ± 0.35 |
| S4 | 6.19 ± 0.36 | 6.23 ± 0.23 |
| S5 | 5.33 ± 0.32 | 5.45 ± 0.12 |
| S6 | 6.11 ± 0.32 | 6.26 ± 0.25 |
| S7 | 2.92 ± 0.12 | 2.87 ± 0.14 |

*4.2. D-value distribution and interpolation*

After the histological studies of the biopsies, here their non-invasive imaging counterpart-$D$-values are computed. The whole tumor volume is covered by 9 DWI slices as described in Section 2.2, which for the original tumor led to 1644 DWI-voxels containing tumor tissue. The $D$-value of each tumor-related voxel is computed to study the $D$-value distribution within the whole tumor. The histogram is shown in Fig. 7(a) in blue. The bin width is set to $1.13 \times 10^{-4}$ $mm^2/s$ according to the Freedman-Diaconis rule (Freedman and Diaconis, 1981), based on the tumor *D*-values before interpolation. The resulting number of bins is $N_{bin}$ = 62. Note that the distribution is asymmetric.

In order to determine the scaling factor *U* for *D*-map interpolation, the *D*-map is interpolated by a scaling factor, increasing from 2 to 60, and the corresponding histogram of the interpolated *D*-values is plotted using the same number of bins $N_{bin}$ = 62. The Kullback-Leibler divergence $Div_{(KL)}$ (Eq. (2)) is computed between the distributions of the original and interpolated *D*-values. $Div_{(KL)}$ tends to converge when *U* is greater than 20 as shown in Fig. 7(a). Thus, a scaling factor of 20 is adopted for the *D*-map interpolation in this study as a trade-off between spatial resolution and computational complexity. For *U* = 20, the final interpolated *D*-map presents a similar histogram as before interpolation (Fig. 7(a)). On a cut of the mid-tumor section shown in Fig. 4(b), although of smoother appearance on interpolated *D*-map as expected, the main image features are conserved.

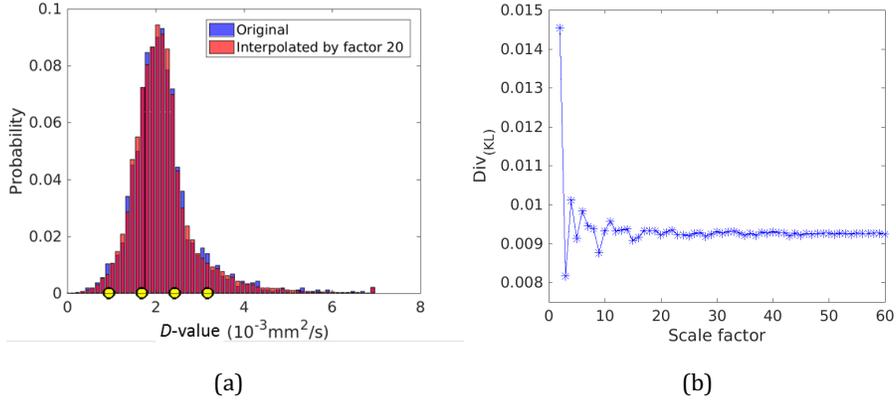

(a)          (b)

**Fig. 7.** *D*-value distribution in the tumor. (a) Original (blue, *Q*) and interpolated (red, *P*) *D*-value histograms of the whole tumor: the *mean* value (±*SD*) of the original *D*-values is $(2.154 ± 0.788) × 10^{-3}\ mm^2/s$, while the value after interpolation by factor 20 is $(2.153 ± 0.765) × 10^{-3}\ mm^2/s$. The yellow points indicate the optimal choice of *D*-values ("optimal D values") when 4 biopsies are proposed to be taken for the studied tumor (see Section 4.3); (b) Kullback-Leibler divergence between the *D*-value distributions before and after the interpolation versus the scaling factor *U* (see Eq. (2)).

### 4.3. Establishment of negative linear relation

The negative correlation between cell density and *D*-value, which was previously investigated from the large tumor blocks in (Yin et al., 2018), can be re-established based on the measures from the virtual biopsies (Fig. 8). Here, the *D*-value is computed from the ROIs on the interpolated *D*-map by Eq. (3). The resulting biopsy data points distribute around the points of the large tumor samples. The angle between the two fitted lines (grey solid and dotted lines in Fig. 8) is 0.001º. Moreover, a Chow test is performed using the *gap* R package (Zhao, 2007). The null hypothesis is that the coefficients of two linear regressions are equal (Chow, 1960). This null hypothesis cannot be rejected with p = 0.996 at a significance level of 0.05, thus there is no significant difference between two linear regressions.

The biopsy data points are next classified into different groups by a model-based clustering using the *mclust* R package (Scrucca et al., 2016). A set of Gaussian mixture models are fitted to the data, estimated by expectation-maximization (EM) algorithm initialized by hierarchical agglomerative clustering. The final mixture model and number of clusters are selected by maximization of Bayesian Information Criterion (BIC). There are in total 5 groups, including 4 groups being closed to the fitted line and a group of points (indicated by orange squares in Fig. 8) surrounding the fat tissue sample that is far from the fitted line. In practice, fat suppression is performed on DWI sequences to suppress the signal from fatty tissue. Therefore, data points of the fat tissue, which produce artifacts should be excluded. Suppressing fat signal leads to an almost perfect linear correlation, with a Pearson coefficient $r$ = -0.95 for large tumor blocks (black solid line) and $r$ = -0.94 for virtual biopsies (black dotted line in Fig. 8). The angle difference between two fitted lines, discarding the fat data points, is 0.002º. There is no significant difference between these two linear regressions (p = 0.903 for Chow test). This means that, excluding fat, the negative linear relationship can be constructed reliably from as few as 4 biopsies that are taken from regions with high, intermediate, and low *D*-values in *D*-map. For the studied tumor, the obtained optimal choice of *D*-values for 4 biopsies are indicated in Fig. 7(a). Our choice of 4 biopsies is guided by the cluster analysis of accessible tissue material. However, the distribution of clusters shown in Fig. 8 suggests that even 3 biopsies are likely to be sufficient if, for example, S2 would be dropped. Generally, the more biopsies are taken, the better the relation between cell density and *D*-value can be represented.

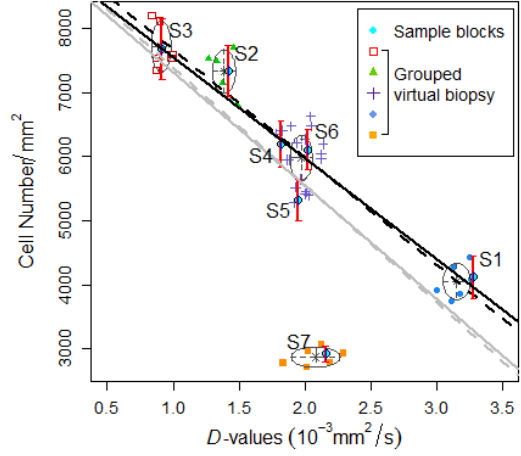

**Fig. 8.** Cell density vs. *D*-value for the large tumor samples (S1-S7, mean value in cyan and standard deviation given by the red bar) and for the five clusters of virtual biopsies (each labeled by a different colored symbol, ellipses indicating the mean vectors and covariances of each component of the Gaussian mixture model associated with the clusters: gray solid line is the fitted line of large tumor samples (Pearson correlation coefficient $r = -0.76$) while gray dotted line is the one for the biopsies ($r = -0.74$); black solid line is the fitted line of large tumor samples excluding fat sample S7 ($r = -0.95$) while black dotted line is the one for the biopsies excluding the data points surrounding sample S7 indicated by orange squares ($r = -0.94$).

*4.4. Cancer cell density and total cell density*

From the histological analysis of tumor samples, it is moreover possible to delineate between cancer and non-cancer cells. By plotting cancer cell density (i.e. cell density for cells marked as cancer cells by cancer cell specific KL1 staining) against total cell density (which includes cancer and non-cancer cells), a positive linear relationship (Pearson correlation coefficient $r = 0.79$) is observed. This relationship is estimated from large tumor samples, while the 95% confidence intervals (CIs) and 95% prediction intervals (PIs) of the cancer cell density are estimated by using *R* statistics (Faraway, 2002) (Fig. 9). The data points of sample S7 of fat tissue are discarded here. For a given value of total cell density, the mean of the cancer cell density will fall in the 95% CIs with 95% probability, and the future observation of the cancer cell density is predicted to be with 95% probability within 95% PIs. Note that most of the biopsy points (solid squares) are within the 95% PIs of the data points of large samples.

Based on the virtual biopsy data, the negative linear relationship of total cell density (in unit *cells/mm²*) vs. *D*-values (in unit $10^{-3}mm^2/s$), as discussed in Section 4.3 (black dotted line in Fig. 8), is expressed as

$$\rho_{2d} = k_{2d}\, D + d_{2d}, \tag{6}$$

where $k_{2d} = -1.684 \times 10^6\ cells \cdot s/mm^4$ and $d_{2d} = 9363\ cells/mm^2$, and the positive linear relationship between the cancer cell density vs. the total cell density is found for $\sim 3000\ cells/mm^2 \leq \rho_{2d}$:

$$\rho_{c,2d} = k_c\, \rho_{2d} - d_c, \tag{7}$$

where, $k_c = 0.43$ and $d_c = 1253\ cells/mm^2$. Knowing the voxel-wise $D$-values within the ROI of the tumor, a total cell density map $P_{total}$ can be generated by calculating the voxel-wise cell density by using Eq. (6) as shown in Fig. 10(a). A cancer cell density map $P_c$ is obtained by estimating the voxel-wise cancer cell density by Eq. (7), and a non-cancer cell density map $P_{nc}$ by subtracting $P_c$ from $P_{total}$, (displayed in Fig. 10(b) and (c) respectively). These maps show a lower cancer cell density compared to the non-cancer cell density, and indicate for all displayed quantities heterogeneity inside the tumor.

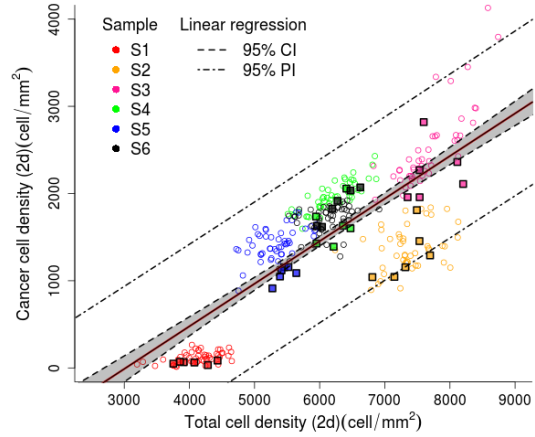

**Fig. 9.** Linear regression of cancer cell density on total cell density: each color represents a tumor sample; each dot represents a tissue slice and each square indicates a virtual biopsy.

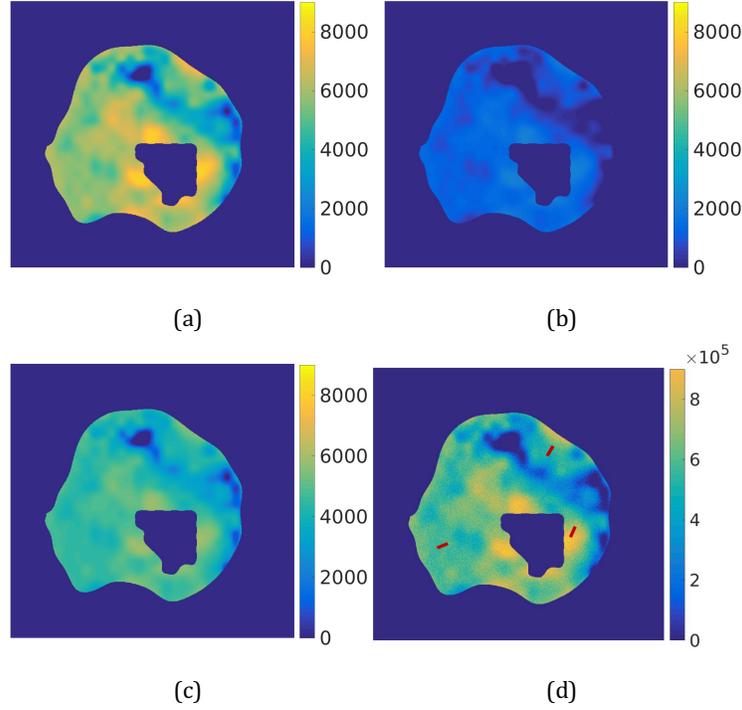

**Fig. 10.** (a) – (c): 2d cell density map (*cells/mm²*) of the middle tumor slice. (a) total cell density map; (b) cancer cell density map; (c) non-cancer cell density map; (d) cross-section of derived 3d synthetic (total) density (*cells/mm³*) map overlaid with three random biopsy regions, the localization being given by the red rectangles showing random access point, needle depths and orientations (red rectangles).

*4.5. Computation of candidate biopsy regions*

Finally, the results for the computation of automated localized biopsies are presented, first in 2d for the middle tumor section and then in 3d for the whole tumor. Table 2 lists the optimal *D*-values of 4 biopsies in 2d and 3d, both of which are estimated from Eq. (4), as well as the final *D*-values of the candidate biopsy regions. As describe in Section 3.4, these candidate biopsy regions are automatically selected as the superpixels that the biopsy needle reaches with the closest values with regard to the optimal *D*-values.

**Table 2**

*D*-values ($10^{-3} mm^2/s$) of the automated localized biopsies in 2d (tumor middle slice) and 3d (whole tumor). The 3d density was calculated from 2d density as explained in Section 3.2.

| ID of biopsy | 1 | 2 | 3 | 4 |
|---|---|---|---|---|
| Optimal *D* (2d) | 0.936 | 1.793 | 2.651 | 3.508 |
| Final *D* (2d) | 0.930 | 1.791 | 2.655 | 3.485 |
| Optimal *D* (3d) | 0.935 | 1.680 | 2.426 | 3.172 |
| Final *D* (3d) | 0.935 | 1.680 | 2.426 | 3.173 |

In 2d, the size of superpixel is set to 6.41 $mm^2$, approximately equal to the size of 18G needle biopsy in literature (Helbich et al., 1998). The partitioned *D*-map and 4 "candidate" biopsy regions are shown in Fig. 11(b) and (c) respectively. "Candidate" denotes regions for which the D values are expected to be close to the "optimal" D values, which are depicted in Fig. 7a for the example of 4 (virtual) biopsies.

In 3d, the size of a supervoxel is set to 4.23 $mm^3$. It is obtained by subdividing the volume (see Fig. 1(b-1)) into small subvolumes using the same concept as for 2d superpixels (see Section 3.4). The final biopsy regions in 3d are obtained as visualized in Fig. 1(b-1), being separated from each other and from the necrotic region in 3d space.

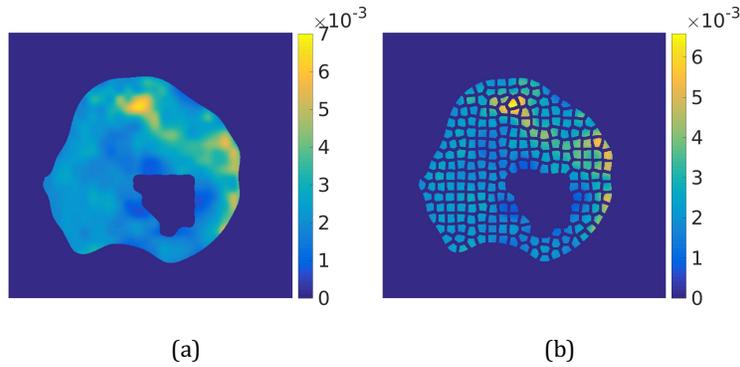

(a)  (b)

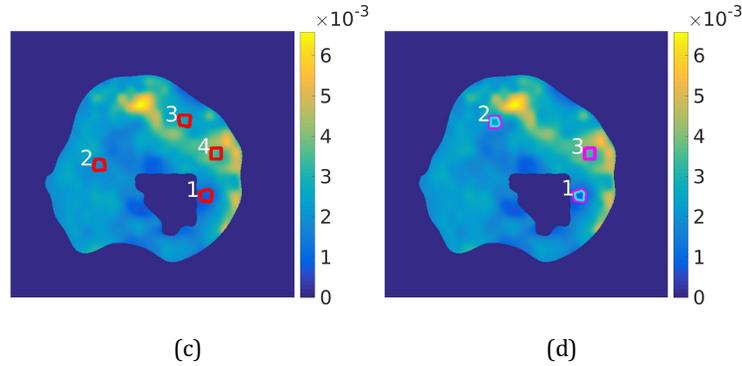

**Fig. 11.** One example of *D*-map ($mm^2/s$) of the middle tumor slice. (a) interpolated *D*-map; (b) partition based on superpixels; (c) Four final candidate biopsy regions; (d) candidate regions of biopsies if the number of biopsies is set to two (cyan) or to three (magenta).

## 4.6. Evaluation of DWI-guided biopsy

### 4.6.1. Synthetic cell density distribution

The real cell density distribution of the whole tumor is unknown as only a part of the tumor could be imaged at the level of histology. In order to assess the DWI-guided and non-guided biopsy procedures, first a synthetic cell density map of the tumor middle slice is generated (Fig. 10(d)) by adding Gaussian white noise $N(0, \sigma_{noise})$ to the estimated total cell density map, which was introduced in Section 4.4. Here, $\sigma_{noise} = 6 \times 10^4$ cells/$mm^3$ is chosen, which is the average standard deviation (SD) of the cell densities computed for the large tumor blocks. Using the same SD for all densities is motivated by the observation that the individual values of the SD at different densities vary only slightly with cell density, not showing any significant multiplicative noise effects.

### 4.6.2. Evaluation of the guided biopsy

If $N_{biopsy}$ needle paths are supposed to be taken during one biopsy intervention, the candidate biopsy regions are chosen by the algorithm discussed in Section 3.4. The algorithm indicates the optimal access point and the biopsy needle paths. The *D*-values of the $N_{biopsy}$ biopsies are computed as the locally averaged values of the precise needle regions (green rectangles in Fig. 12). The corresponding cell densities of the $N_{biopsy}$ biopsies are computed as the local cell density gathered by the needles from the synthetic cell density map. For the example of $N_{biopsy} = 3$, three access points are found at the tumor border for which the needles can reach the candidate biopsy regions. Fig. 12 shows the result of needle paths from one of the three access points.

To investigate the influence of the number of biopsies, the guided biopsy procedure is simulated for $N_{biopsy}$ = {2, 3, 4}. For each case, the correlation between the *D*-values and the cell densities from the final found needle paths is re-established (see Fig. 13(a)-(c)). The number of fitted lines in each subfigure is equal to the number of optimal access points found by the algorithm.

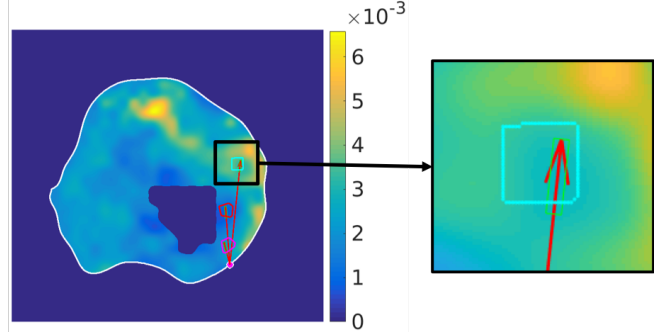

**Fig. 12.** Guided biopsy result on *D*-map ($mm^2/s$): candidate regions in red, pink and cyan have the representative low, middle and high *D*-values, respectively. The pink point at the tumor border is the access point found by the proposed algorithm. The red arrows indicate the needle paths (depth and orientation inside the tumor). The zoom-in picture on the right shows the precise region in green (width = 0.5 mm; length = 2.5 mm) where the needle could take the tissue material from the candidate region (a local superpixel marked in cyan).

Once a strictly monotonic relationship is obtained, which here is linear, the *D*-value for each voxel of tumor volume on *D*-map can be converted into the number of cells (population size) within that voxel. The total cell population size of the tumor is the sum of the cell population sizes of each voxel belonging to the tumor. Fig. 13(d) shows the estimated tumor cell population size for different $N_{biopsy}$. The estimation accuracy is high under DWI-guidance, comparing the estimated values with the reference number of tumor cell load 2.57×$10^{10}$ (pink horizontal line in Fig. 13(d)) which is estimated from the large tumor blocks (Yin et. al., 2018). This value is used as reference as no 'ground-truth/real' number of tumor cells is possible to be obtained. There is little difference between the results of different number of biopsies. When $N_{biopsy}$ = 4, the accuracy is a bit higher than for $N_{biopsy}$ = 2, 3 when the mean-value is considered. The reliability of the guided biopsy procedure is also reflected in the distributions of sample mean $\bar{\rho}$ and standard deviation *s* for different number of needle paths $N_{biopsy}$ = {2, 3, 4} (Fig. 14a, b), which is compared to non-guided biopsies below.

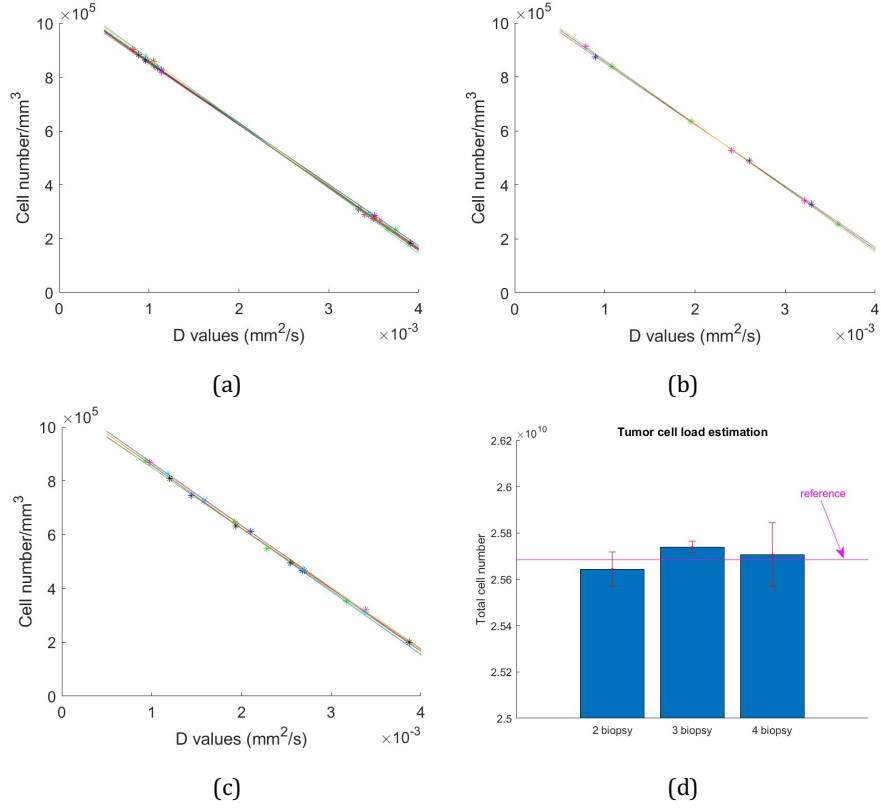

**Fig. 13.** Re-established linear relationship between the *D*-value and cell density. Each colored dot represents the value of one needle path. (a) $N_{biopsy}$ = 2; (b) $N_{biopsy}$ = 3; (c) $N_{biopsy}$ = 4; (d) total cell load estimation.

*4.7. Comparison among guided, non-guided with constraints, and random biopsies*

*4.7.1. Sample mean and standard deviation*

When performing biopsy intervention with $N_{biopsy}$ needle paths, let $\rho_j$ indicate the cell density of tumor tissue material gathered by the needle along the $j^{th}$ needle path on the synthetic cell density map for $j$ = 1, ..., $N_{biopsy}$. The sample mean of cell density is

$$\bar{\rho} = \frac{1}{N_{biopsy}} \sum_{j=1}^{N_{biopsy}} \rho_j, \tag{8}$$

and the estimated sample standard deviation is

$$s = \sqrt{\frac{1}{N_{biopsy}-1} \sum_{j=1}^{N_{biopsy}} (\rho_j - \bar{\rho})^2}. \tag{9}$$

The cell density mean $\mu_\rho$ and standard deviation $\sigma_\rho$ computed over the tumor region on the synthetic cell density map (Section 4.6.1) are used as references in the following discussion.

*4.7.2. Non-guided biopsy*

Here, non-guided biopsy is defined as biopsies performed without DWI guidance. Two cases, with and without constraints (constraints as defined in section 3.5) are taken into account.

- With constraints: for one single biopsy intervention (a set of needle paths punched through one access point) all sets of the possible biopsy access points, needle depths and orientations are simulated. Fig. 14(c) and (d) show the distributions of sample mean $\bar{\rho}$ and standard deviation $s$ for different number of needle paths $N_{biopsy}$ = {2, 3, 4}. When $N_{biopsy}$ increases, $\bar{\rho}$ seems almost unchanged while the peak of distribution of $s$ is closer to the reference $\sigma_\rho$. This biopsy procedure without DWI guidance has very low probability to represent the reference $\mu_\rho$ and $\sigma_\rho$. In other words, the tumor cell load ($Volume_{tumor} \times \mu_\rho$) and heterogeneity (standard deviation $\sigma_\rho$) cannot be reliably estimated.
- Without constraints (random biopsy): the biopsy is taken through random access point positions and orientations as illustrated in Fig. 10(d). For different needle biopsy numbers $N_{biopsy}$ = {2, 3, 4}, the random procedure is repeated for 20,000 times. Fig. 14(e) and (f) show the distributions of $\bar{\rho}$ and $s$. When $N_{biopsy}$ increases, the random biopsies are more probable to represent the reference $\mu_\rho$ and $\sigma_\rho$ if compared to the biopsies of non-guided with constraints. However, the probability to catch the tumor heterogeneity ($\sigma_\rho$) is still low when $N_{biopsy}$ = 4.

*4.7.3. Comparison among the different biopsy procedures*

Three cases of simulations, the guided, and non-guided biopsies with and without constraints are compared as shown in Fig. 14. The guided biopsy is by far more likely to represent the reference tumor cell density $\mu_\rho$ (Fig. 14(a)) and the chance to significantly deviate from the reference values is very small as the density distribution is sharply peaked. For non-guided biopsies, either with (Fig. 14(c)) or without (Fig. 14(e)) constraints, the probability to miss the reference $\mu_\rho$ is large as the distribution of sample mean $\bar{\rho}$ is broad with long tails. For non-guided biopsies, the presence of constraints induced asymmetry in the distributions of $\bar{\rho}$. When more needle paths are taken during one biopsy intervention, the effect of $N_{biopsy}$ on $\bar{\rho}$ is low in all three cases (Fig. 14(a), (c) and (e)).

Furthermore a measure to evaluate the spread of density values is investigated that is accessed for one biopsy intervention in the three cases. For example, in Fig. 14(b) the $N_{biopsy}$ = 2 (3, 4) means that the tissue sample consists of tissue material from 2 (3, 4) needle paths from the same access point. In order to construct a reliable relation between *D*-value and cell density, the different needle paths should sample tissue from regions that cover a large range of *D*-values (Section 4.2), which for $N_{biopsy}$ = 2 means sampling from regions with high and with low *D*-values that indicate small and large cell density. As a consequence, the sample standard deviation *s* between two density values $\rho_1$ and $\rho_2$ should be large, which is the case denoted by blue bars in Fig. 14(b). Hence, the fact that the sample standard deviation is higher for the guided biopsies is by design since each sample should reach well-separated *D*-values. As $N_{biopsy}$ increases, *s* naturally decreases which is why with increasing $N_{biopsy}$, *s* approaches the reference $\sigma_\rho$ (Fig. 14(b)). On the contrary, for non-guided biopsies, whether constrained (Fig. 14(d)) or not (Fig. 14(f)), *s* for $N_{biopsy}$ = 2 is small as random sampling has a high chance to sample in regions with comparable *D*-values. Moreover, *s* increases with $N_{biopsy}$ as increasing the number of needle paths increases the chance of sampling in regions of different *D*-values, hence different cell densities (Fig. 14(d) and (f)). With increasing of $N_{biopsy}$ in all three cases, *s* moves towards the reference $\sigma_\rho$ (pink vertical line in Fig. 14(b), (d) and (f)).

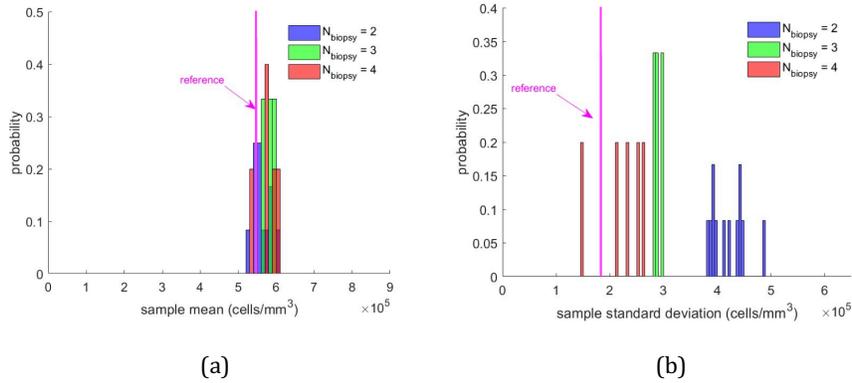

(a)             (b)

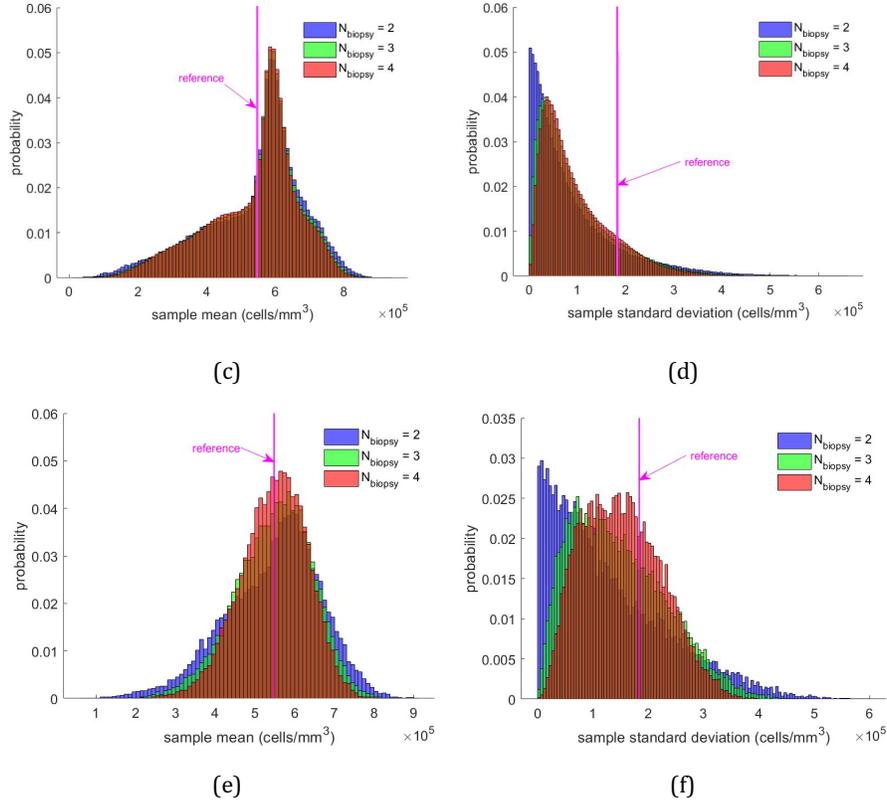

**Fig. 14.** Distributions of estimated sample mean $\bar{\rho}$ and standard deviation $s$ for DWI-guided biopsies ((a), (b)), and non-guided biopsies with ((c), (d)) and without ((e), (f)) constraints for $N_{biopsy}$ = {2, 3, 4}. The reference lines in pink denote the reference mean $\mu_\rho$ and standard deviation $\sigma_\rho$ computed from the synthetic cell density map, respectively. For guided biopsy procedure whose distributions of $\bar{\rho}$ and $s$ shown in (a) and (b), the numbers of biopsy interventions (sets of virtual biopsies paths) for $N_{biopsy}$ = {2, 3, 4} were 12, 3, and 5 respectively. For non-guided biopsies without constraints ((c) and (d)), 277,000 biopsy interventions were simulated for $N_{biopsy}$ = 2, and more than 2,000,000 for $N_{biopsy}$ = {3, 4}. For random biopsies ((e) and (f)), 20,000 virtual biopsy interventions were taken for $N_{biopsy}$ = {2, 3, 4}.

## 5. Discussion and conclusion

In this study, a novel way of estimating tumor cell load and heterogeneity from combining DWI with needle biopsies is proposed. More specifically, a protocol is established to infer from the tumor DWI derived *D*-map the most informative regions where to take biopsies in the 3d tumor volume.

In this work virtual biopsies are defined based on large tumor samples from a NSCLC patient-resected-tumor that were previously analyzed in detail. The representativeness of the biopsies is studied by comparing cell density distribution and *D*-values distribution between these biopsies and the large tumor samples in (Table 1, Fig. 8 and Fig. 9).

Based on the observation that tumor cell density is linearly correlated to the *D*-value, a new protocol for biopsy-based tumor evaluation is proposed (described in Section 1). The protocol is however expected to apply also in case of a non-linear correlation as long as it is strictly monotonic. Among these steps, the localization of biopsies is key. In the clinics, biopsies are currently suggested to be performed in tumor regions of high cell density (based on a homogeneous region of low *D*-value in DWI), or high tracer uptake in PET, or high flow region in DCE-MRI. Here a superpixel-based approach to partition the tumor *D*-map by grouping pixels with similar features is proposed, thus dividing the tumor *D*-map volume into sub-regions. By selecting the most representative *D*-values from these superpixels (as defined by Eq. (4)), roughly regions of low, high and intermediate *D*-values, the final regions from which the biopsies should be taken are determined. The optimal needle paths are probed by simulation of the biopsy procedure taking into account clinical constraints. After histological analysis of the biopsies, the relationship between cell density and *D*-values is re-constructed, and the tumor cellularity and heterogeneity obtained. From the analysis of clusters classifying the tumor biopsy samples into four groups with regard to their *D*-value and cell density, spanning from small to large *D*-values (hence well representing the tumor heterogeneity) and excluding fat, four biopsy samples have been found that permit to reconstruct well the relation between cell density and *D*-value compared to previously obtained values by studying large tumor tissue blocks (Yin et. al., 2018). The results presented in Fig. 8 and Fig. 13(d) suggest that even two biopsy samples might be sufficient, if the relation between density and *D*-value is linear as for the NSCLC tumor in this study. However, our study results recommend taking at least three-four samples capturing the range of *D*-values (one small, intermediate, large *D*-value) to obtain a warning in case linearity is not fulfilled, which might give rise to take additional biopsy samples. Of course, the more biopsy samples are taken, the more robust and accurate the relation can be established but there is a pay-off between robustness/accuracy on one hand and patient side effects on the other hand. Currently, Univ. Clinics of Heidelberg routinely takes three biopsy samples for NSCLC patients for diagnostic purposes. The approach developed in this work might represent the basis for new clinical algorithms in case biopsy material is or can be taken for diagnostics. Taking 3-4 biopsies from vital and non-necrotic tumor tissue following our proposed sampling algorithm permits to compute tumor cellularity and tumor cell load (number of cells in the tumor), which could provide valuable information to clinicians in diagnosis and therapy planning or surveillance. For example, tumor cellularity might directly enter into the calculation of the amounts of cytotoxic drugs given to patients. So far, this information is not considered in therapeutic treatment protocols, which means that patients with a high number of tumor cells may receive insufficient doses. In contrast, patients with low cellularity may suffer from unnecessary adverse effects because they receive too high drug doses. More important, tumor heterogeneity is one critical parameter for the development of chemo-resistance (Dagogo-Jack and Shaw, 2018); high tumor heterogeneity may favor radiotherapy, which permits dose-painting to adapt to tumor heterogeneity (Alfonso et. al., 2014). Indeed, the knowledge provided by our algorithm would be essential to identify patients with higher degrees of tumor heterogeneity. These cancer patients might have to be monitored more carefully, since the probability for a relapse might be higher under selective pressure as e.g. from drug administration.

The conceptual approach presented in this paper for NSCLC may have to undergo further studies and discussions prior to application. For example, cellularity and heterogeneity do significantly differ between tumor types (e.g. lung cancer and liver cancer) or even within distinct tumor entity (e.g. adenocarcinoma and squamous-cell carcinoma of the lung). This could indicate that the number of biopsies must be individually defined for different cancers and subtypes. Thus, it would recommendable in future work to verify the results of the current study on a variety of tumor types, for example by analyzing in how far the heterogeneity caught by biopsies taken as proposed by the algorithm sufficiently reflects the entire tumor in other tumor types and subtypes. On the other hand, an ethical discussion is needed if higher numbers of biopsies are acceptable and appropriate for patients. The later point is of special interest since for some tumor entities needle biopsies may lead to unwanted tumor cell seeding along the needle channel (Robertson and Baxter, 2011). However, for the tumor type of this current study, our study suggests changing the biopsy design, but not the number of biopsies compared to the current protocol at Univ. Clinics of Heidelberg.

As discussed in the introduction, biopsies can be taken under different imaging modalities. However, in general, biopsies cannot be easily taken in closed MR scanners, although the direct MRI-guided procedures could be feasible with specialized tools (needles, markers, and devices) designed with low magnetism materials that are acceptable for MR systems (Weiss et al., 2008). The co-registration of the $D$-map with images from open MR, CT or ultrasound might be necessary.

Finally, this work removes a strong limitation of our previous work (Yin et. al., 2018) as here cellularity is accessed patient-specific prior to therapy only using biopsies and non-invasive DWI. This avoids the need of large tumor samples requiring surgical intervention and/or the existence of tumor-specific calibration curves linking *D*-value and tumor cell density that would have to be determined across many tumors and sampled from many patients.

**Appendix A.**

The *D*-value considered here is thought to probe intercellular diffusion (Yin et. al., 2018).

To smooth the *D*-value map requires an assumption of how *D* depends on the cell density $\rho$. For demonstration purpose consider a one-dimensional medium with average cell density $\rho_i = \rho(x_i)$ associated with voxel *i*. If one assumes that the local cell density changes sufficiently smoothly from one voxel *i* to its neighbor voxel $j = i \pm 1$, then the simplest spatial density profile between $\rho_i$ and $\rho_{i+1}$ would be a linear profile, hence the density at any intermediate point in the space interval $x_i < x < x_{i+1}$ would be $\rho(x) = \rho(x_i) + (\rho(x_{i+1}) - \rho(x_i))\frac{x-x_i}{x_{i+1}-x_i}$. *D* is associated with the water diffusion constant and assumed to reflect the density, with high densities relating to small *D* and vice versa, which might be reflected by the simple phenomenological relationships $D \propto \rho_{max} - \rho$ in the range $\rho \in [0, \rho_{max}]$. The relationship leads to $\rho = \rho_{max} A(1 - \frac{D}{D_0})$ with *A* being some proportionality constant, which is the relation found identified in ref. (Yi et. al., 2018, see also Fig. 1(c-2)). From Fig. 1(c-2), $\rho \approx 1.125 \times 10^6 - 2.5 \times 10^5 D$, with $\rho$ in $cells/mm^3$ and *D* in $10^{-3} mm^2/s$. Hence, the theoretical value at *D = 0* is $\rho(D = 0) \approx 1.125 \times 10^6 cells/mm^3 = 1.125 \times 10^{-3} cells/\mu m^3$, meaning that the average volume associated with one cell at *D = 0* would be $v = \frac{V}{N} = \rho^{-1} = 0.89 \times 10^3 \mu m^3$ (*V* being a tissue volume, *N* the number of cells in that volume). This corresponds to a cuboidal space of edge length $\sim 9.6\ \mu m$ associated with one cell. As this corresponds approximately to a typical cell diameter, at *D = 0* the cells are largely space-filling hence $A \approx 1$. This line of argument is also supported by the histological images at the maximal density observed in the data (Fig. A.1(a)). The self-diffusion coefficient of water at *40°*C is $D_w \sim 3.5 \times 10^{-3} mm^2/s$ which is the value at $\rho \approx 2.5 \times 10^5 cells/mm^3$, where cells do not limit water diffusion anymore (the *D*-value at $\rho = 0$ is $D \approx 4.5 \times 10^{-3} mm^2/s$ but given the diffusion constant of water at body temperature is smaller, the linear relation does probably not apply anymore at values of *D > D_w*). The largest *D*-values observed were at $\sim 3 \times 10^{-3} mm^2/s$, where the cell density was small, so that diffusion of water would be expected to be largely unconstrained (Fig. A.1(b)).

If on the other hand, the tissue would change its viscosity, the diffusion constant might be related inversely to the viscosity, $D \propto \frac{1}{\eta}$, which would lead to interpolation of $1/D$ rather than $D$. Several ways of printing $D$ and $\rho$ or their inverse were tested and it was found that the relation depicted in Fig. 1(c-2) gave the best agreement with the data.

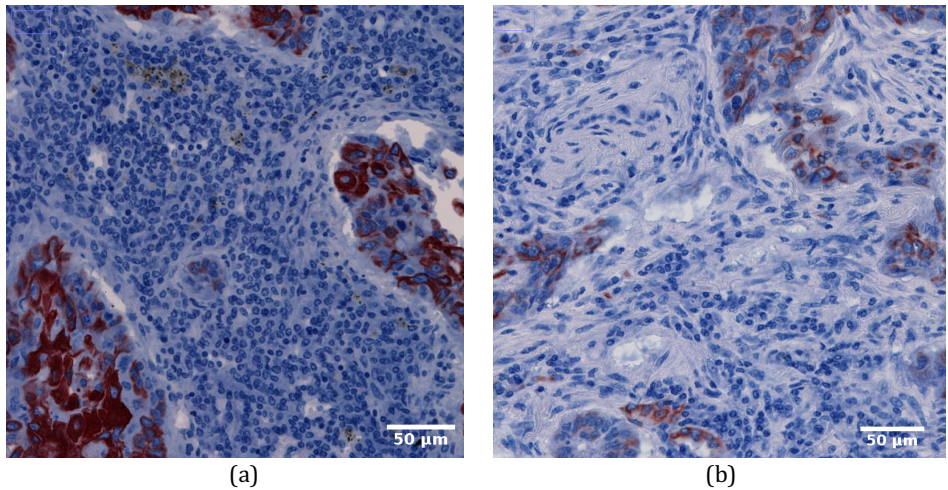

(a)  (b)

**Fig. A.1.** Exemplary images from histology showing the density of cells $\rho$ (in cells/$mm^3$) in the tumor at (a) $D(\rho = 8 \times 10^5 cells/mm^3) = 10^{-3}\ mm^3/s$, (b) $D(\rho = 4.5 \times 10^5 cells/mm^3) = 3 \times 10^{-3}\ mm^3/s$. The carcinoma cells are stained in brown (KL1). The tumor in (a) is already very dense even though perhaps not completely compact. The immune cells located between the cancer islands are smaller than $10\mu m$ but the cancer cells are bigger. In (b), there are large spaces between the cells, so diffusion of water molecules almost unconstrained.

**Acknowledgment**

This work was supported by the BMBF-Projects Cancer-Sys/LungSys through the German Federal Ministry of Education and Research under Grants FKZ 0316042H and 0316042B. Y. Yin further acknowledges support from EU-NOTOX, ANR iFlow and D. Drasdo from ANR-INVADE, PRT-K16 (Inst. Natl. du Cancer), ANR iLite and BMBF-LiSym.